\documentstyle[twocolumn,aps]{revtex}

\begin{document}

\title{Overcoming the Wall in the Semiclassical  Baker's Map } 

\author { L. Kaplan \thanks{kaplan@physics.harvard.edu} }
\address{Department of Physics, Harvard University\\
Cambridge, Massachusetts 02138\\}

\author { E. J. Heller \thanks{heller@physics.harvard.edu} }
\address{Department of Physics, Harvard University and\\
Harvard-Smithsonian Center for Astrophysics\\
Cambridge, Massachusetts 02138\\}
\date{\today}
\maketitle

\begin{abstract}

A major barrier in semiclassical calculations is the sheer
number of terms that contribute as time increases;
 for classically chaotic dynamics, 
the proliferation is exponential.
We have been able to overcome this ``exponential wall''  for the baker's map,
using an analogy with spin-chain partition functions.
The semiclassical sum is contracted so that only of order $N\times T^{3/2}$
operations are needed for an $N$-state
system evolved for $T$ time steps. This is typically less than the
computational load for quantum  propagation, $T\times N^2$.
This method enables us to obtain semiclassical results up to the
Heisenberg time, 
which  in our example  would have required $10^{90}$ terms if we were to
evaluate the sum exactly.   This calculation in turn
provides new insight as to the accuracy of the semiclassical
approximation at long times.
The  semiclassical result is  
often correct;  its breakdown is nonuniform. 

\end {abstract}

  \vskip .3cm



In previous work it was shown that semiclassical time correlators
(and autocorrelation functions in particular)
for chaotic dynamical systems can give accurate answers
past the log time, where the semiclassical approximation had
been thought to break down\cite{log}.  The log time is the time it takes
a given cell of size $h^D$ (for $D$ degrees of freedom)  to visit essentially 
all other cells.  
Theoretical justifications showing why the log time is not always expected
 to be the end of semiclassical accuracy were given\cite{breakdown}.
  Essentially, 
it is not the number of terms in the semiclassical sum that should
cause a breakdown, but rather more standard concerns about the validity 
of the stationary phase approximation for each term,
which in turn depends on the areas of enclosed regions
in phase space.  Often, an {\it algebraic} breakdown is
expected, and, perhaps counterintuitively, strong chaos can 
{\it improve} the semiclassical expressions by creating healthy areas
(greater than $\hbar$) in the phase plane.

Just beyond the log time in a semiclassical propagation, 
another problem  emerges: the   numerical
burden of summing over all  the classically independent 
paths leading from the   initial position and returning 
to it.  This  ``exponential wall'' of proliferating contributing
orbits has brought many semiclassical calculations to a halt, well
before they were actually breaking down. This was the case 
for the  baker's map\cite{baker} and the stadium\cite{log},
both of which are completely chaotic.  The wall exists for any strongly chaotic
system, since the number of contributing terms in a semiclassical
propagator grows exponentially with time.  (Similar problems besiege the
energy domain trace formula\cite{gutz}, where  exponential growth occurs
with the  action or length of the contributing orbits.)  
 The exponential wall is doubly 
frustrating because it both limits the utility of semiclassical methods as a 
tool for calculations and hides from view the answer to an important question:
when  and how do time-domain  semiclassical methods  break down?  

Since a small change in the classical mechanics does not change the quantum 
mechanics very much, but can change the long-time classical mechanics
drastically, intuition suggests that the details of individual classical
returning orbits are not important, especially at longer times.  
A detailed, explicit semiclassical sum over all returning 
orbits is thus not really necessary. 
The  information in all the returning trajectories at long times is
much larger than the information contained in the eigenstates,
again suggesting that some summarizing of the
classical mechanics would do little harm to the semiclassical sum. 
 
The work of Dittes, Doron, and Smilansky \cite{dds}, 
which approximated  the baker's map transfer operator using
a finite  Fourier transform, successfully achieved such a summarizing of the
classical mechanics in a novel way, and provided the first glimpses of 
breakdown well past the log time.  We return at the end to some of the issues
they  raised about the accuracy of the long-time semiclassical 
propagation. 

The classical baker's map is a famous paradigm of chaos and mixing.  It was
quantized first by Balazs and Voros\cite{bv}, and has received steady
attention since then. Such paradigm systems are one of the great strengths
of modern chaos theory, because of the qualitative properties that survive
translation into many  situations.  

The classical baker's map is an area-preserving, discrete map
of the unit square onto itself defined by horizontally expanding and
vertically compressing the left half of the square so as to map it onto the
bottom half, and similarly taking the right half into the top half. The map is 
hyperbolic with constant exponent $\lambda = \ln(2)$ and is smooth everywhere
except for a set of measure zero (the ``cutting region" $q = 1/2$). For our
purposes, a symbolic representation is most convenient: phase space is the set
of infinite sequences of zeroes and ones, with a decimal point inserted (the
digit following the decimal point identifies the phase space point as being
in the left or right half of the square),
and the dynamics is given by shifting the decimal point
one place to the right at each time step. This description makes it easy to
find periodic and homoclinic orbits.
(We note that every string
$\gamma$ of length $P$ is naturally associated with a periodic point of period
$P$, the point with symbolic representation
$...\gamma\gamma\gamma.\gamma\gamma\gamma...\,$.
Alternatively, if the wavepacket
is centered on periodic point $...\alpha\alpha\alpha.\alpha\alpha\alpha...\,$,
it may be useful to think of the string $\beta$ as standing for a homoclinic
excursion, with trajectory
$...\alpha\alpha\alpha\beta\alpha\alpha\alpha...\,$.)

A quantization preserving the symmetries of the classical system 
\cite{saraceno} is obtained
by choosing an even integer $N$ ($N = 1/h$) and defining a $q$-basis of states
lying at $q=(n+1/2)/N$, $n=0 \, .. \, N-1$, and a $p$-basis similarly (this
quantization corresponds to imposing doubly antiperiodic boundary conditions
on the unit square). 
Planck's constant $h$ (a dimensionless number, since we have defined
phase space area to be unity) serves here as the grid spacing.
The dynamics
is most easily described in a mixed representation, with the leftmost $N/2$
$q$-states mapping (through a discrete Fourier transform) to the bottom $N/2$
$p$-states, and similarly for the remaining part of the Hilbert space. A
relative phase between the two blocks in the propagator matrix is undetermined,
and leads to a one-parameter family of quantizations. Ignoring this phase, in
configuration space the propagator has the matrix form
$B = \left[ F^{-1}_N \right]
 \left[ \begin{array}{cc} F_{N/2} & 0 \\ 0 & F_{N/2} \end{array} \right]\,,$
where $F_N$ is the discrete Fourier transform on N sites.

A quantity of interest is the quantum autocorrelation function
$A^\Psi(T) = <\Psi|B^T|\Psi>$, where $\Psi$ is some wavepacket
(e.g., a coherent state). This function, and its semiclassical analogue
$A^\Psi_{SC}(T)$, when Fourier transformed, provide information on the
quantum and semiclassical spectra and eigenfunctions of the system. Although
the statistical properties of the autocorrelation function and its Fourier
transform, as well as the divergence between these quantities and their
semiclassical approximations for large times (low frequencies), are
quite interesting objects of study, in this letter we focus on the actual
evaluation
of $A^\Psi_{SC}(T)$ for particular values of $\Psi$ and $T$. In contrast with
the quantum autocorrelator, which can be computed in polynomial time
(by matrix multiplication), the semiclassical autocorrelator can be computed
exactly only in exponential time, because of the exponential proliferation of
contributing classical paths
for large $T$. We propose a method of approximating this sum that should also
be applicable to other systems in which a symbolic dynamics can be used to
classify the classical orbits.

We work in the coherent state representation, and begin with a Gaussian 
centered at $P_{cent}=(q_{cent},p_{cent})$
in the unit square, with width $\sigma$ in
the $q$ direction. The $2^T$ classes of
possible trajectories are labelled by strings $\beta$ of $T$ binary digits.
Each class  of trajectories corresponds to a unique symbolic
history; geometrically, a class corresponds to a stretched (in $q$) and 
compressed (in $p$) piece of the original disc in phase space representing the 
initial Gaussian.  
The total overlap at time $T$ can be written as   

\begin{equation}
\label{sum}
A^\Psi_{SC}(T) =
a(T) \sum_{strings \, \beta} e^{-f(\beta,P_{cent},\sigma,T)}
e^{i g(\beta,P_{cent},\sigma,T)}\,,
\end{equation}
where $a(T)$ falls off exponentially for large $T$, compensating for the
increase in the number of orbits to be summed over, and $f$ and $g$
are quadratic polynomials in $q_{cent}$, $p_{cent}$, and $\beta$.
One can think of the initial Gaussian  as freely spreading
along the $q$-direction and contracting
along the $p$-direction for time $T$, then being cut into pieces
and put back into the unit square according to the map.  Each piece
is part of the stretched Gaussian; thus the universal 
prefactor $a(T)$. Each term in the sum (\ref{sum}) corresponds to the 
overlap of the initial Gaussian with one of the stretched pieces. 
The pieces  have aquired a phase $g$ that is specific to the history 
of each piece; each piece is attenuated depending on its q-position in the
stretched Gaussian and also ``misses'' the $p-$center of the original
Gaussian by some amount, and thus the corresponding overlap is attenuated by
a total factor $\exp(-f)$.
Specifically, the exponential suppression of a contribution depends 
on the location of the
periodic point $P_\beta$ associated with string $\beta$, relative to the
wavepacket center $P_{cent}$.
In terms of the symbolic dynamics, roughly speaking $P_\beta$ falls inside the
wavepacket if the first $\log_2(N)/2$ bits of $\beta$ match the binary
expansion of $q_{cent}$ and the last $\log_2(N)/2$ bits, taken in reverse
order, match the expansion of $p_{cent}$. (This is true for a circular
wavepacket ($\sigma=\sqrt\hbar$); for an elliptical wavepacket elongated
in the $q$ or $p$
directions, the relative importance of the beginning and end of the string 
will be different, but the total number of bits which are essentially fixed,
i.e. $\log_2(N)$, is unchanged.) 

The  suppression exponent $f(\beta)$ and phase $g(\beta)$ are both quadratic
forms in the coordinates of the point $P_\beta$, and hence quadratic in the
digits of the finite binary string $\beta$. This permits the
autocorrelator (\ref{sum}) to be
written as a partition function for a finite-length spin chain with an external 
potential and two-body interactions between the spin-1/2 particles. The
role of temperature is played by $1/N$.

The ``external potential" ({\it i.e.} the piece of $-f+ig$ linear in the
digts of $\beta$) turns out to be important only for the two ends of
the chain, in the large-$T$ (long-chain) approximation. Similarly
the two-body interactions in the real part $f(\beta)$ are
significant only for bits within $\sim \log_2(N)$ of either edge. {\em Thus,
for large $T$, the vast majority of bits in the string $\beta$ enter into the
expression (\ref{sum}) only through the two-body contribution to the phase
$g(\beta)$.}
This contribution, however, has the form
$\exp(i \pi N\sum_{i>j}({1 \over 2})^{i-j}\beta_i\beta_j)$,
where $\beta_i$ (= 0 or 1) is the
$i$-th bit of string $\beta$. {\em Therefore, the relevant
``spin-spin interactions"
are local,
with scale $\log_2(N)$.} We can use this fact to approximate the sum by
considering blocks of bits, with length of order $\log_2(N)$. These blocks
then have only nearest-neighbor interactions, and the sum can be done
just as for an Ising chain. The approximation can be improved in a controlled
way by increasing the size of blocks used. The result is that to any arbitrary
desired accuracy, the semiclassical autocorrelator can be computed in
polynomial time for large $T$, just as is the case for the quantum 
autocorrelator.

Ignoring factors of order unity, the expected phase error
that results from omitting
interactions between bits separated by more than the cutoff of $C$ sites
goes as $N ({1 \over 2})^C T^{1/2}$. 
(The $T^{1/2}$ factor arises from an incoherent sum over $T$ terms.)
Because phase errors from different
orbits add incoherently, just as the total phases themselves do, this is
also expected to be the fractional size of the error in the sum
$A_{SC}^{\Psi}(T)$. Thus, the number
of operations required to evaluate the autocorrelator up to some large time
$T$ with error $\epsilon$ behaves as $N T^{3/2} / \epsilon$, compared
with $2^T$ operations for evaluating this sum exactly.
Typically, one
wants to take the calculation out to a time of order the Heisenberg time
(the inverse of the typical eigenvalue spacing),
which is $N$ for the baker's map. Thus,
the computational load is of order $N^{5/2} / \epsilon$, compared with
$2^N$ for obtaining the full semiclassical result. 
Thus substantial savings can be achieved for
large $N$ (where the semiclassical approximation may be
expected to be relevant), with minimal loss in precision.
The exact {\it quantum}
propagation of a state vector for the Heisenberg time
requires order $N^3$ operations ($N^2$ multiplies for each
of  $N$ time steps), more than the contracted semiclassical sum.

We will present numerical data for a wavepacket centered on the period-2
orbit given by $\alpha=01$. The wavepacket is centered on $(1/3,2/3)$,
well away from any cuts, so the semiclassical approximation is very good
even for moderate values of $N$. (In any case, the behavior of the baker's
map near the cut is the one aspect of this system
which is certainly non-generic).
The value of $N$ used in obtaining the
data in Fig.~1 is $N=226$. (Note that a value equal to 2 times a prime number
has been chosen to guarantee ``generic" behavior, free from number theoretic
anomalies. In particular, values of $N$ equal to a power of 2 should be avoided
for the standard baker's map -- such values lead to anomalously large
recurrences due to coherent interference effects.) 
The semiclassical autocorrelator is seen to
follow the quantum result very closely until about $T=N/4$, at which time
higher order
quantum effects become noticeable. Even for $T$ comparable to the Heisenberg
time, though, some of the behavior of the autocorrelator function is still
seen in the semiclassical calculation. We must note, however, that
there is a definite upper bound to the
accuracy of the semiclassical approximation
in the time domain coming from the non-unitarity of semiclassical physics
\cite{dds}. If we had taken the calculation out to several times the
Heisenberg time, the exponentially growing eigenstates would begin to dominate
the autocorrelator, even for a wavepacket located well away from the caustic
in the map.
 
Looking in the energy domain, we find that some spectral peak locations
and heights are reproduced very well, while others are reproduced only
poorly and yet
others not at all. This is consistent with the results obtained in \cite{dds};
our interest however is in understanding the semiclassical behavior of
individual states (coherent states or eigenstates, for example), because
too much information is lost when concentrating only on the trace
of the propagator. In particular, though we do not discuss these issues in
this letter, the methods of \cite{dds} can easily and, in our opinion
fruitfully, be extended to study the properties of eigenstates which are
well approximated by the semiclassical physics as well as those which are not.
Doing this for the baker's map and other systems should shed light on the
nature of the breakdown of the semiclassical approximation at large times.

Although the present paper deals only with the standard baker's map, the
approach described here should be generalizable to other maps. The extension
to generalized baker's maps, with multiple vertical strips of possibly unequal
width, seems to be most straightforward. However, systems for which a
symbolic description requires a grammar should also be treatable by these
methods. Special care will be necessary in applying our approach to
systems in which the Lyapunov exponent is either vanishing or very small
in some
regions of phase space. For such systems, it will not always be the case that
the ``interactions" between symbols widely separated in time are small, and
such phase space regions will therefore need to be treated separately from
the ``hard chaotic" regions.


We believe the work we report here may also shed light on the energy-domain 
Gutzwiller trace formula.
The   trace formula attempts to give  individual eigenvalues in terms of
purely classical  periodic orbits of all lengths.  It is 
derived by stationary-phase Fourier transform from the
semiclassical  time-domain
Green function that we study here, and stationary phase trace over all 
coordinates.  
To understand the  notorious convergence difficulties of the Gutzwiller
trace formula,  one should see how long the  semiclassical time-domain
Green function remains valid.  If it 
fails to reach the Heisenberg time, then blame for failure of the 
trace formula to predict individual eigenenergies could be laid at the feet
of the diffraction and tunneling which degrade the time-domain semiclassical
Green function.  Further, the methods we use to contract the proliferating
homoclinic orbits may suggest new ways to re-sum the trace 
formula, a subject which has recently seen much attention in the form of cycle
expansions and other tools which effectively (and tantalizingly in the light
of the time-domain results given here and earlier) re-order
orbits according to time of arrival\cite{cycle}.

Our results  extend the work of Dittes,  Doron, and Smilansky.
  We note that  their approach, involving a finite Fourier transform of the 
transfer operator,
 is completely different from the spin-chain sum used here.  
Their conclusions about the breakdown were not very optimistic, but  they
studied mainly  the trace of the propagator, rather than 
its action on particular states.  
Dittes {\it et. al.} also studied the quasienergy spectrum and  attributed 
the breakdown of the trace to certain quasienergy eigenvalues
which had modulus greater
than unity (unitarity necessitates unit modulus of the exact 
results).  They cast doubt on the previously claimed
algebraic improvement of accuracy with decreasing $\hbar$\cite{breakdown},
because the  bad eigenvalues have a damaging effect on the trace.  However,
the trace in fact cloaks a nonuniform
breakdown in phase space of the type that is presaged in earlier work with the
baker's map\cite{baker}. There it was noted that nonstationary states initiated 
near the ``cut'' in the map suffer diffraction after being cut
and degrade much more rapidly than states initiated elsewhere. 
In the present study, 
we  see evidence that the breakdown is indeed very nonuniform, striking some 
initial nonstationary states more strongly than others, and affecting
some semiclassical eigenstates much more than others.

The present work   reinforces
 the hope that approximate evaluation of semiclassical sums will 
continue to develop as a tool, making many more semiclassical calculations
possible.  The approximation given here, which
depends on  symbolic dynamics and  draws on analogies
with spin-chain partition functions, can be   {\it faster} than
doing the quantum mechanics.  This is a rare trait
 in semiclassical approximations to chaotic systems, and
 represents something of a watershed for semiclassical
long time dynamics. 
(We  note that the Dittes,  Doron, and Smilansky method
requires a matrix representation larger in dimension than the original 
quantum problem, thus their method is necessarily
 somewhat slower than doing a full quantum calculation.)
These results may hint at possibilities for further progress in work with 
periodic orbits and the  Gutzwiller trace formula\cite{gutz}.

  There are strong indications 
that other maps and dynamical systems will yield to similar simplifications.

\vskip 0.1in

Fig. (1a). Semiclassical vs. quantum autocorrelator for initial
circular wavepacket centered on $(1/3,2/3)$ periodic point (phases not
shown).

Fig. (1b). Spectrum obtained by Fourier transforming time data shown above.

\section{Acknowledgments}
 
We acknowledge important discussions with S. Tomsovic.  This 
research was supported by the National Science Foundation under grant number 
CHE-9014555.



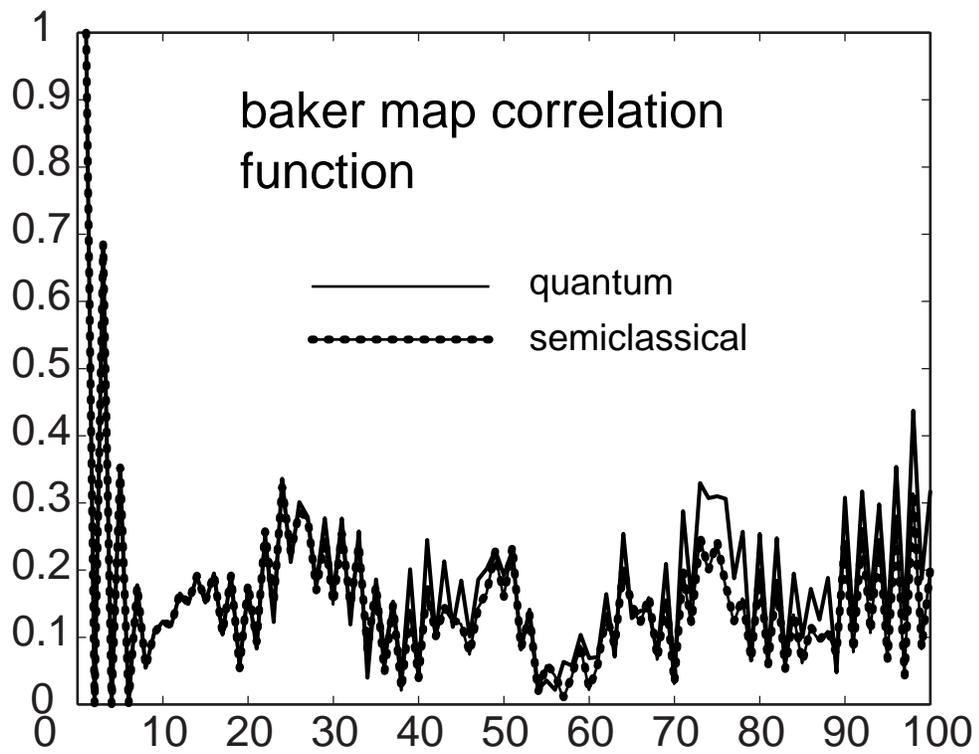
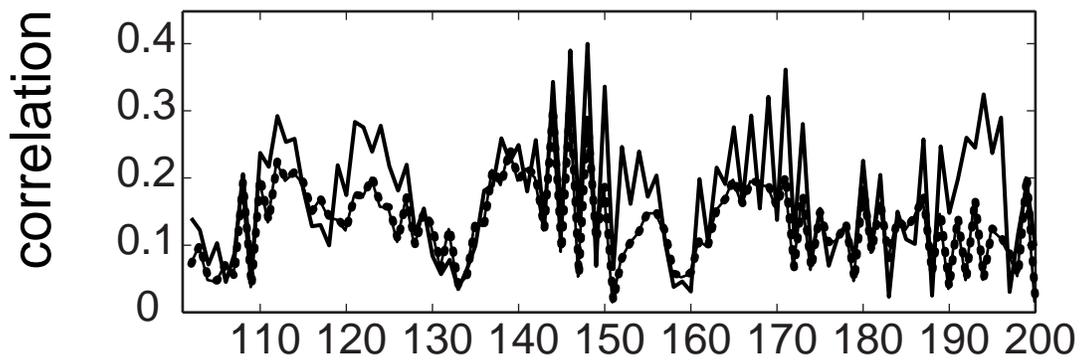
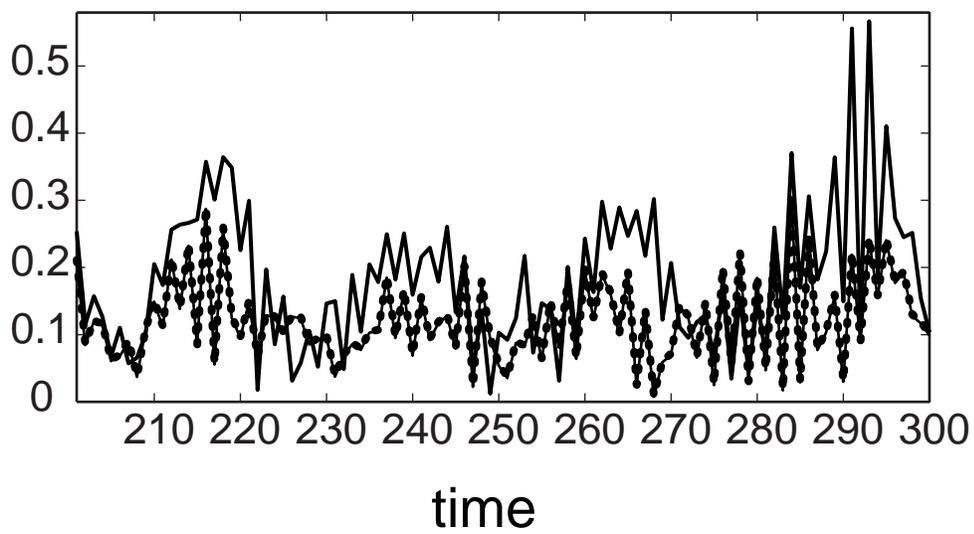
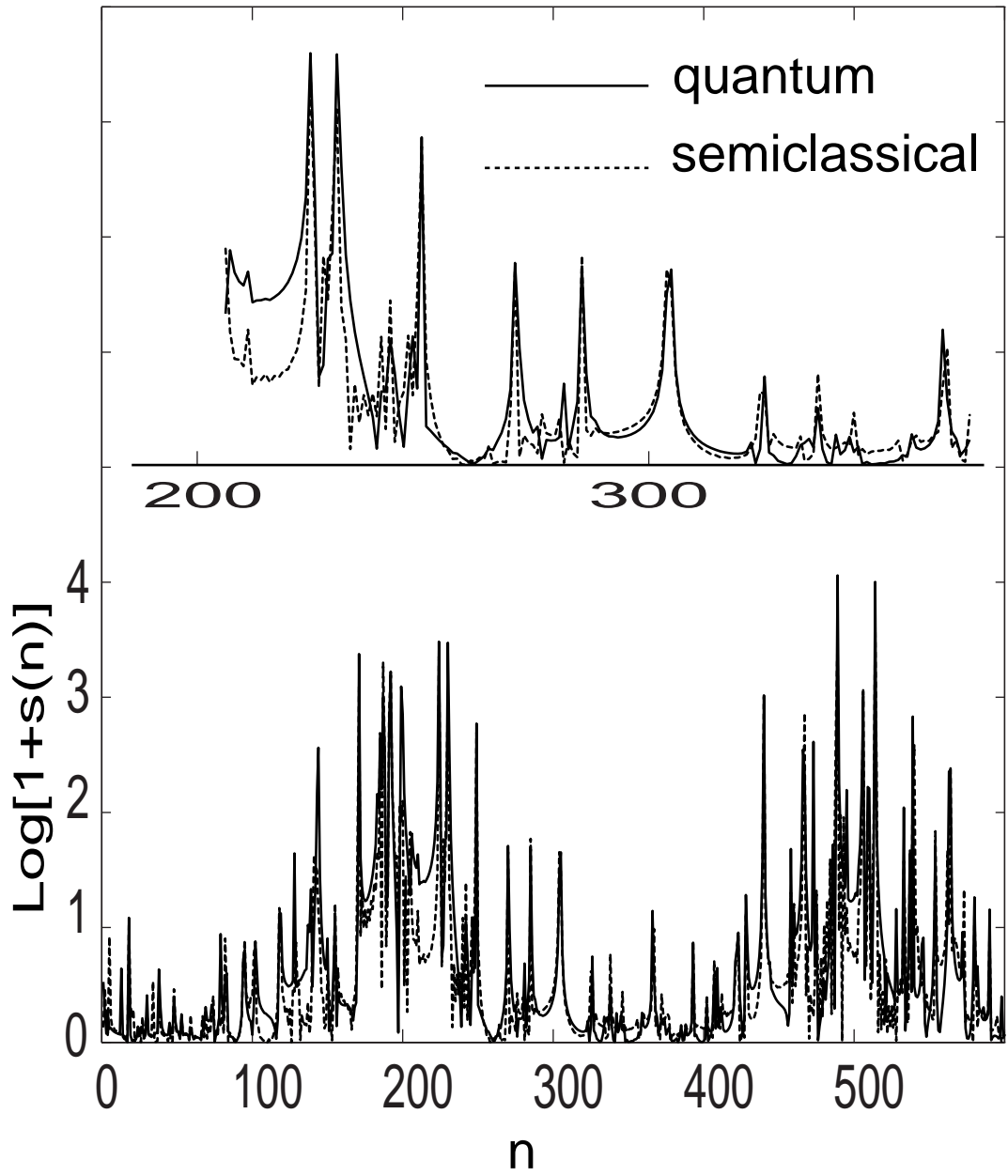
\end{document}